\documentclass[10pt,aps,prb,longbibliography,nobibnotes,twocolumn,groupedaddress]{revtex4-1}
\usepackage[utf8]{inputenc}
\usepackage{placeins}
\usepackage{tabularx}
\usepackage{amsmath}
\usepackage{amsfonts}
\usepackage[colorlinks,citecolor=blue,urlcolor=blue,bookmarks=false,hypertexnames=true]{hyperref}
\usepackage{amssymb}
\usepackage{graphicx}
\usepackage{float}
\usepackage{threeparttable}
\usepackage{subcaption}
\usepackage[dvipsnames]{xcolor}
\usepackage[shortlabels]{enumitem}
\usepackage{dcolumn} 
\usepackage{braket}
\newcolumntype{d}[1]{D{.}{.}{#1}}
\begin{document}
\def\ie{{\it i.e.\/}}
\def\eg{{\it e.g.\/}}
\def\etal{{\it et al.\/}}
\def\cm{{cm$^{-1}$}}
\def\kJmol{kJ$\,$mol}
\def\Eh{$E_\mathrm{h}$}
\def\mEh{m$E_\mathrm{h}$}
\def\uEh{$\mu E_\mathrm{h}$}
\def\a0{$a_0$}
\def\bea{\begin{eqnarray}}
\def\eea{\end{eqnarray}}

\title{DLPNO-MP2 for Periodic Systems using Megacell Embedding}
\author{Andrew Zhu}
\affiliation{University of Oxford, South Parks Road, Oxford, OX1 3QZ, UK}
\author{Arman Nejad}
\affiliation{University of Oxford, South Parks Road, Oxford, OX1 3QZ, UK}
\author{Poramas Komonvasee}
\affiliation{University of Oxford, South Parks Road, Oxford, OX1 3QZ, UK}
\author{Kesha Sorathia}
\affiliation{University of Oxford, South Parks Road, Oxford, OX1 3QZ, UK}
\author{David P. Tew}
\affiliation{University of Oxford, South Parks Road, Oxford, OX1 3QZ, UK}

\date{\today}

\begin{abstract}

We present a domain-based local pair natural orbital M{\o}ller--Plesset second order perturbation theory (DLPNO-MP2) for periodic systems, working within an LCAO formalism within the Tubromole program package. This approach, Megacell-DLPNO-MP2, embeds a supercell correlation treatment within a megacell and does not involve periodic image summation for the Coulomb integrals. Working in a basis of well-localised Wannier functions, periodicity is instead imposed through rigorous translational symmetry of Hamiltonian integrals and wavefunction parameters. The accuracy of the method is validated through comparison with a complementary periodic DLPNO-MP2 method that employs Born--von K{\'a}rm{\'a}n boundary conditions, described in paper I of this series. The PNO approximations are shown to be equivalent in the two approaches and entirely consistent with molecular DLPNO-MP2 calculations. The Megacell-DLPNO-MP2 method displays sub-linear scaling with respect to supercell size at the asymptotic limit and example calculations are presented with up to 15000 basis functions in the correlation treatment.


\end{abstract}
\maketitle

\section{Introduction}
\label{sec:Intro}
Electronic structure studies of materials have been largely dominated by the use of density functional theory\cite{dft1,dft2} (DFT), due to its excellent cost-to-accuracy ratio. 
However, the uncertainties associated with density functional approximations are difficult to characterise and interest in employing wavefunction-based methods has grown considerably
\cite{cryscor,usvyat1,usvyat2,gtoccberkelbach,gruneisthermodycc,hiratamp2,HirataCCperiodic,goldzaksosmp2,scuseriamp2,lnoccberkelbach,nagasemp2ri,boothfciqmcsolid,gruneisccreview,cscmp2berkelbach,schaferquarticmp2,gruneisboothnatural,gruneismp2cclih,michaelidesmgoco,berkelbachmgoco,qcvsqmcusvyat,usvyatcomgowater}.
Post Hartree--Fock (HF) formalisms such as M{\o}ller--Plesset (MP) perturbation theory and coupled cluster (CC) theory provide a hierarchy of  systematically improvable methods for treating many-body correlation,  and several periodic implementations of these approaches are now available \cite{cryscor,usvyat1,usvyat2,gtoccberkelbach,gruneisthermodycc,hiratamp2,HirataCCperiodic,goldzaksosmp2,scuseriamp2,lnoccberkelbach,nagasemp2ri}. The major barrier to their wider use within computational materials science is their expense. Canonical second-order MP perturbation theory (MP2) scales as $\mathcal{O}(N^5)$ with system size, whilst canonical coupled cluster with singles, doubles and perturbative triples (CCSD(T)) scales as $\mathcal{O}(N^7)$. Practical calculations encounter severe memory and CPU bottlenecks whenever larger or more complex unit cells are employed, or when increasing the number of simulated unit cells to approach the thermodynamic limit.

Domain-based pair natural orbital local correlation\cite{neese2009efficient,neese2013} (DLPNO) theory offers a solution to this problem. DLPNO theory achieves near linear scaling of computational effort with system size with only modest loss in accuracy by replacing Hamiltonian integrals and excitation amplitudes with low-rank approximations that exploit the inherent locality of electron correlation in insulators. For each electron pair, a bespoke set of virtuals are constructed to efficiently capture the pair correlation energies. In the context of molecular electronic structure theory, the DLPNO approach has been successfully applied to perturbative\cite{wernerpnolmp2,Schmitz01092013,neesemp2,tewhattigpnomp2,hattigtewpnomp2mp3f12} and coupled cluster formalisms\cite{dlpnoccsdt,dlpnoccsdtimproved,wernerpnolccsd,wernerpnolccsdtf12,hattigpnoccsdtaccuracy,hattigpnoccsdt,neesedlpnoccsdtbench,neesedlpnoccsdnewnearlinear,linearscaling1}, to explicitly correlated theory\cite{wernerpnolmp2f12,pnoccsdpnomp2tew,wernerpnolccsdf12,tewhattigpnomp2,tewf12principal}, to multireference methods\cite{pnomscaspt2,yanaimrpno,neesenevpt2pno} and to excited states\cite{hattigexcited,hattigexcitedlinear,neeseexcited,valeev-pno-excite}, greatly extending the range of applicability of these approaches for computing energies and properties of molecular systems.


A number of alternative local correlation schemes for periodic systems have previously been proposed. In particular, the pioneering work within the Cryscor\cite{cryscor1,cryscor,cryscorosv,usvyatlmp2f12,usvyat1,usvyat2,usvyatlih} package leverages projected atomic orbitals\cite{Pulay} (PAOs), and orbital specific virutals\cite{osvs} (OSVs) to obtain a low-scaling MP2 implementation for non-conducting systems. More recently,  Ye and Berkelbach\cite{lnoccberkelbach} have adapted K\'allay's local natural orbital (LNOs) approach\cite{kallaylno,kallayccsdt,kallaylnoccsdtoptim,kallaylnoccsdtoptim2} to extended systems, at the CCSD and CCSD(T) levels of theory.  These methods have been applied successfully to obtain properties such as lattice constants, cohesive energies\cite{lnoccberkelbach,usvyatlmp2cohesive,usvyat2} and adsorption energies\cite{usvyatcomgowater,berkelbachmgoco,usvyatargonmgo,usvyatlmp2al2o3} for surface interactions. PNO theory has notable advantages over these alternative schemes. The pair-specific nature of PNOs affords a much greater degree of compression of the correlation space than PAOs or OSVs, where domains for distant pairs are the same size as for close pairs. PNO theory also provides a route to the efficient computation of excited states and response properties\cite{hattigexcitedlinear,hattigexcited,neeseexcited,valeev-pno-excite}.

In this work, we present a periodic, real-space, local MP2 method based on DLPNO theory. Our implementation leverages the existing molecular DLPNO-MP2 code in the Turbomole program package and is made possible due to the periodic Hartree--Fock implementation in the \verb:riper:\cite{riper1,riper2,riper3,riper4,riper5} module. We exploit the translational invariance of orbitals, amplitudes and integrals to obtain an effective sub-linear scaling of computational effort with respect to unit cell number. Our approach employs a hybrid PAO-OSV-PNO framework, using PAOs and OSVs as intermediates to improve efficiency for PNO evaluation. 
 A DLPNO-MP2 implementation provides the foundation for a DLPNO-CCSD(T) treatment for periodic systems.

We are concurrently pursuing two complementary approaches to simulating periodic systems. One where Born--von K{\'a}rm{\'a}n (BvK) boundary conditions are applied to the correlation treatment, leading to electron repulsion integrals involving lattice summations over the periodic images, and one that embeds a supercell correlation treatment in a megacell, where periodicity is imposed through translational invariance and integrals do not involve lattice summations. The BvK scheme is presented in Paper I of this series\cite{bvk}. This contribution presents the second approach, which we call Megacell-DLPNO-MP2. We demonstrate the performance of the Megacell-DLPNO-MP2 method for a set of one-, two- and three-dimensional insulating or semi-conducting systems, by monitoring the convergence to the thermodynamic and canonical limits with supercell size and PNO thresholds, respectively. The BvK-DLPNO-MP2 method is used to validate the accuracy of Megacell-DLPNO-MP2 in later sections, a detailed analysis and comparison of the two methods will be the focus of future work.

\section{Theory}
\label{sec:mp2}
\subsection{Local MP2 for Periodic Systems}

Within the linear combination of atomic orbitals (LCAO) framework, under Born--von K{\'a}rm{\'a}n boundary conditions, real space basis functions, known as BvK AOs, are expressed as
\begin{equation}\label{eq:bvkao}
	\ket{\mu_{\mathbf{l}}} =\sum_{\mathbf{L}}^{\infty} \ket{\mathring{\mu}_{\mathbf{l+L}}} .
\end{equation}
Here, $\mu$ labels the AO within a unit cell and $\mathbf{l}$ is a lattice vector index labeling the unit cell in the BvK `supercell'. The BvK AO has the periodicity of the BvK supercell since it is the infinite sum over all periodic images of the AO $\mathring{\mu}$, where $\mathbf{L}$ is the vector index labeling the supercell,
spanning the infinite crystal. Periodic Hartree--Fock calculations are performed in the basis of 
Bloch AOs, which are eigenfunctions of the crystal momentum operator with eigenvalue $\mathbf{k}$
\begin{equation}\label{blochao}
	\ket{\mu_{\mathbf{k}}} =\frac{1}{\sqrt{N}}\sum_{\mathbf{l}}^N  e^{i \mathbf{k}\cdot \mathbf{l}}  \ket{\mu_{\mathbf{l}}},
\end{equation}
where $N$ is the number of unit cells within the BvK supercell. The BvK boundary condition imposes discretizations on the momenta $\mathbf{k}$, which is equivalent to defining a supercell size in real space. The conversion between BvK AOs and Bloch AOs is in this case a generalized discrete Fourier transform (FT) and our choice of normalization is such that the FT matrix is unitary.

The overlap and Fock matrices are block diagonal in the Bloch AO basis, due to translational symmetry, and the Hartree--Fock orbitals and eigenvalues satisfy
\begin{align}
\mathbf{F}_\mathbf{k}^{\mathbf k} \mathbf{C}_\mathbf{k}^\mathbf{k} &= \mathbf{S}_\mathbf{k}^\mathbf{k} \mathbf{C}_\mathbf{k}^\mathbf{k} {\mathcal E}_\mathbf{k}
\end{align}
where
\begin{align}
\langle \mu_\mathbf{k} \vert \nu_{\mathbf k'} \rangle &= \delta_\mathbf{k k'} S_{\mu_\mathbf{k}}^{\nu_\mathbf{k'}}  \\
\langle \mu_\mathbf{k} \vert \hat F \vert \nu_{\mathbf k'} \rangle &= \delta_\mathbf{k k'} F_{\mu_\mathbf{k}}^{\nu_\mathbf{k'}}. 
\end{align}
Solution of the HF equations yields the canonical crystal orbitals (COs) $\vert p_\mathbf{k}\rangle$, also referred to as Bloch functions, and orbital eigenvalues $\epsilon_{p_{\mathbf{k}}}$,
\begin{equation}\label{Bloch-1}
	\ket{p_{\mathbf{k}}} =\sum_{\mu} \ket{\mu_{\mathbf{k}}}C_{\mu_\mathbf{k}}^{p_\mathbf{k}}.
\end{equation}
Under a spin-free MP2 formalism, the amplitude and energy equations in the canonical basis are
\begin{align}
0 &= g^{i_{\mathbf{k}_1}j_{\mathbf{k}_2}}_{a_{\mathbf{k}_3} b_{\mathbf{k}_4}} +
(\epsilon_{a_{\mathbf{k}_3}} + \epsilon_{b_{\mathbf{k}_4}} - \epsilon_{i_{\mathbf{k}_1}} - \epsilon_{j_{\mathbf{k}_2}})
t^{i_{\mathbf{k}_1}j_{\mathbf{k}_2}}_{a_{\mathbf{k}_3} b_{\mathbf{k}_4}}, \\
E_\mathrm{corr} &= \frac{1}{N} 
\sum_{i_{{\mathbf k}_1} j_{{\mathbf k}_2} a_{{\mathbf k}_3} b_{{\mathbf k}_4}}
(2g_{i_{\mathbf{k}_1}j_{\mathbf{k}_2}}^{a_{\mathbf{k}_3} b_{\mathbf{k}_4}} - g_{i_{\mathbf{k}_1}j_{\mathbf{k}_2}}^{b_{\mathbf{k}_4} a_{\mathbf{k}_3}})
t^{i_{\mathbf{k}_1}j_{\mathbf{k}_2}}_{a_{\mathbf{k}_3} b_{\mathbf{k}_4}},
\end{align}
where $g^{i_{\mathbf{k}_1}j_{\mathbf{k}_2}}_{a_{\mathbf{k}_3} b_{\mathbf{k}_4}}
= \langle i_{\mathbf{k}_1} j_{\mathbf{k}_2} \vert r_{12}^{-1} \vert a_{\mathbf{k}_3} b_{\mathbf{k}_4}\rangle$ 
are the Bloch electron repulsion integrals (ERIs) and $E_\mathrm{corr}$ is the second-order correlation energy per unit cell. Occupied orbitals are denoted by $i,j$ and virtuals by $a,b$.

The canonical COs are delocalized throughout the entire crystal and are not suitable for local correlation approximations. By applying an inverse FT, one can generate Wannier Functions (WFs)\cite{Wannier} $\vert p_{\mathbf{l}}\rangle$, which are a real space representation of the orbitals, each centered on a unit cell given by lattice vector $\mathbf{l}$,
\begin{equation}
	\ket{p_{\mathbf{l}}}=\frac{1}{\sqrt{N}} \sum_{\mathbf{k}}^N e^{-i \mathbf{k}\cdot\mathbf{l}} \ket{p_{\mathbf{k}}}.
\end{equation}
Wannier functions obey translational symmetry, whereby each function $\ket{p_{\mathbf{l}}}$ is a translational copy of the corresponding function in the reference cell $\ket{p_{\mathbf{0}}}$, noting that the translation may wrap around the periodic supercell boundary. Wannier functions can be further localized by rotating the Bloch functions $\vert p_{\mathbf{k}}\rangle$ at each $\mathbf{k}$-point among themselves prior to Fourier transformation. We have recently introduced a procedure for obtaining well-localized Wannier functions by optimizing a fourth-order Pipek--Mezey\cite{PM} metric using atomic charges from Bloch intrinsic atomic orbitals (IAOs)\cite{knizia2013,zhuwannier}, which we
summarize in Section \ref{sec:wf}. 

In the Wannier basis, the MP2 amplitude and energy equations become
\begin{align} \label{eq:mp2reswannier}
0 =& g^{i_{\mathbf l_1} j_{\mathbf l_2}}_{a_{\mathbf l_3} b_{\mathbf l_4}} +
\sum_{c_{\mathbf l_5}} (f_{a_{\mathbf l_3}}^{c_{\mathbf l_5}} t^{ i_{\mathbf l_1} j_{\mathbf l_2}}_{c_{\mathbf l_5} b_{\mathbf l_4}} + f_{b_{\mathbf l_2}}^{c_{\mathbf l_5}} t^{ i_{\mathbf l_1} j_{\mathbf l_2}}_{a_{\mathbf l_3} c_{\mathbf l_5}}) \nonumber \\
&- \sum_{k_{\mathbf l_5}} (t^{ k_{\mathbf l_5} j_{\mathbf l_2}}_{a_{\mathbf l_3} b_{\mathbf l_4}} f_{k_{\mathbf l_5}}^{i_{\mathbf l_1}} + t^{ i_{\mathbf l_1} k_{\mathbf l_5}}_{a_{\mathbf l_3} b_{\mathbf l_4}} f_{k_{\mathbf l_5}}^{j_{\mathbf l_2}}),
\end{align}
\begin{align}\label{eq:emp2wannier}
E_\mathrm{corr} &= \frac{2}{1+\delta_{i_{{\mathbf 0}}j_{{\mathbf l}}}} \sum_{i_{{\mathbf 0}} \le j_{{\mathbf l}_2}}\sum_{a_{{\mathbf l}_3} b_{{\mathbf l}_4}}
(2g_{i_{\mathbf{0}}j_{\mathbf{l}_2}}^{a_{\mathbf{l}_3} b_{\mathbf{l}_4}} - g_{i_{\mathbf{0}}j_{\mathbf{l}_2}}^{b_{\mathbf{l}_4} a_{\mathbf{l}_3}})
t^{i_{\mathbf{0}}j_{\mathbf{l}_2}}_{a_{\mathbf{l}_3} b_{\mathbf{l}_4}},
\end{align}
where the electron repulsion integrals (ERIs) in the BvK AO basis are evaluated as
\begin{align}
g_{\mu_{\mathbf{l}_1} \nu_{\mathbf{l}_2}}^{\kappa_{\mathbf{l}_3} \lambda_{\mathbf{l}_4}} = 
\sum_{\mathbf{L}_2\mathbf{L}_3\mathbf{L}_4}^\infty \langle \mathring{\mu}_{\mathbf{l}_1} \mathring{\nu}_{\mathbf{l}_2+\mathbf{L}_2} \vert r_{12}^{-1} \vert \mathring{\kappa}_{\mathbf{l}_3+\mathbf{L}_3} \mathring{\lambda}_{\mathbf{l}_4+\mathbf{L}_4} \rangle.
\label{eq:eri}
\end{align}
The Fock matrix elements are
\begin{align}
f_{p_{\mathbf{l}}}^{q_{\mathbf{m}}} &= \frac{1}{N}\sum_{\mathbf k} e^{-i \mathbf{k}\cdot\mathbf{l}} f_{p_\mathbf{k}}^{q_\mathbf{k}} e^{i \mathbf{k}\cdot\mathbf{m}} \\
\mathbf{f}^{\mathbf k}_{\mathbf k} &= {\mathbf{U}^{\mathbf k}_{\mathbf k}}^\dagger \mathbf{F}^{\mathbf k}_{\mathbf k} {\mathbf{U}^{\mathbf k}_{\mathbf k}},
\end{align}
where $\mathbf{U^k_k}$ are the unitary matrices obtained from the Bloch IAO localization procedure. 
The restricted summation in Eq.~\ref{eq:emp2wannier} counts each pair interaction once and only includes pairs where at least one orbital is in the reference unit cell, which is at lattice displacement $\mathbf 0$. We adopt the convention that the reference cell is at the centre of the supercell and use an odd number of $\mathbf{k}$-points in each dimension.

Having transformed to the Wannier basis, local approximations can be applied to reduce the computational costs. Electrons in distant orbitals $j_{\mathbf{l}}$ have negligible correlation with 
electrons in orbital $i_{\mathbf{0}}$ of the reference unit cell and the number of pairs 
$i_{\mathbf{0}}j_{\mathbf{l}}$ with pair energy greater than $\epsilon$ tends to a constant as the size of the supercell is increased. In the PNO approach to local correlation, a model pair density is used to determine an
$\mathcal{O}(1)$ set of pair-specific localized virtual orbitals $\{\tilde a_{i_{\mathbf{0}}j_{\mathbf{l}}}\}$ 
adapted to describe the correlation of each occupied pair $i_{\mathbf{0}}j_{\mathbf{l}}$. The error incurred due to discarding virtuals is proportional to $\sqrt{{\mathcal T}_\text{PNO}}$, where ${\mathcal T}_\text{PNO}$ is the PNO occupation number threshold that defines the PNO subspace\cite{keshaextrpolate,keshadamyan}. In the DLPNO approach, the PNOs are expanded in a pair-specific domain of PAOs $\{\tilde{\mu}_{\mathbf{m}} \}$

\begin{align}
\vert \tilde a_{i_{\mathbf{0}}j_{\mathbf{l}}} \rangle = \sum_{\tilde{\mu}_{\mathbf{m}} \in \mathcal{D}_{i_{\mathbf{0}}j_{\mathbf{l}}}} \vert \tilde{\mu}_{\mathbf{m}} \rangle \tilde{C}_{\tilde{\mu}_{\mathbf{m}}}^{\tilde{a}_{i_{\mathbf{0}}j_{\mathbf{l}}}}
\end{align}
The energy and amplitude equations for periodic DLPNO-MP2 are analogous to those for the molecular case,
\begin{align}\label{eq:pnomp2res}
0 =&\,\,
g^{i_{\mathbf 0} j_{\mathbf l}}_{\tilde a \tilde b} + 
(\varepsilon_{\tilde a} + \varepsilon_{\tilde b}) t^{i_{\mathbf 0} j_{\mathbf l}}_{\tilde a \tilde b} \nonumber \\
& - \sum_{k_{\mathbf m}} \sum_{\tilde c \tilde d \in [ k_{{\mathbf m}}j_{{\mathbf l}} ] }
S_{\tilde a, i_{\mathbf 0} j_{\mathbf l}}^{\tilde c, k_{\mathbf m} j_{\mathbf{l}}}
S_{\tilde b, i_{\mathbf 0} j_{\mathbf l}}^{\tilde d, k_{\mathbf m} j_{\mathbf{l}}}
t^{ k_{\mathbf m} j_{\mathbf l}}_{\tilde c \tilde d} f_{k_{\mathbf m}}^{i_{\mathbf 0}}  \nonumber \\
& - \sum_{k_{\mathbf m}} \sum_{\tilde c \tilde d \in [ i_{{\mathbf 0}}k_{{\mathbf m}} ] }
S_{\tilde a, i_{\mathbf 0} j_{\mathbf l}}^{\tilde c, i_{\mathbf 0} k_{\mathbf{m}}}
S_{\tilde b, i_{\mathbf 0} j_{\mathbf l}}^{\tilde d, i_{\mathbf 0} k_{\mathbf{m}}}
t^{ i_{\mathbf 0} k_{\mathbf m}}_{\tilde c \tilde d} f_{k_{\mathbf m}}^{j_{\mathbf l}} \,,\\
E_\mathrm{corr}=& \,\,\frac{2}{1+\delta_{i_{{\mathbf 0}}j_{{\mathbf l}}}}\sum_{i_{{\mathbf 0}}\le j_{{\mathbf l}}} \sum_{\tilde a \tilde b \in [ i_{{\mathbf 0}}j_{{\mathbf l}} ] }
 (2g_{i_{\mathbf{0}}j_{\mathbf{l}}}^{\tilde a \tilde b} - g_{i_{\mathbf{0}}j_{\mathbf{l}}}^{\tilde b \tilde a})
t^{i_{\mathbf{0}}j_{\mathbf{l}}}_{\tilde a \tilde b} \,,
\label{eq:emp2}
\end{align}
where $\tilde a \tilde b \in [ i_{\mathbf{0}} j_{\mathbf{l}} ]$ and $S_{\tilde a, i_{\mathbf{0}} j_{\mathbf{l}}}^{\tilde c, k_{\mathbf{m}} j_{\mathbf{l}}}$ is the overlap between PNOs for pair $i_{\mathbf 0} j_{\mathbf l}$ and pair $k_{\mathbf m} j_{\mathbf{l}}$.  For convenience, the PNOs are rotated to diagonalise the block of the Fock matrix that they span, with diagonal elements $\varepsilon_{\tilde{a}}$. Since the WFs are translationally symmetric, the PAO, OSV and PNO domains are also rigorously translationally symmetric. The amplitudes, overlaps and Fock matrix elements spanning the entire supercell can therefore be generated from the translationally unique set of objects where at least one index resides in the reference cell. For example, $f_{k_{\mathbf m}}^{j_{\mathbf l}}=f_{k_{\mathbf m - \mathbf{l}}}^{j_{\mathbf 0}}$ and $t^{ k_{\mathbf m} j_{\mathbf l}}_{\tilde c \tilde d}=t^{ k_{\mathbf 0} j_{\mathbf l - \mathbf{m}}}_{\tilde c \tilde d} $, where it is understood that modulo arithmetic is applied to the lattice vectors.

A linear scaling algorithm for the correlation energy can be constructed in the same way as for the molecular case by forming the PNOs through a PAO to OSV to PNO subspace compression cascade and using local density fitting for the ERIs. In contrast to the molecular case, however, the BvK boundary conditions require that all functions are periodic over the supercell. Consequently, overlap, Fock and ERI integrals involve summing each function's contribution within all periodic images of the supercell, up to the thermodynamic limit, as shown in Eq.\ref{eq:bvkao}. This lattice summation presents a significant complication. In particular, evaluation of the periodic Coulomb integrals becomes more challenging due to the long-range nature of electrostatic interactions\cite{ashcroft1976solid}, which requires careful accounting of the charge and higher multipole contributions to ensure absolutely convergent lattice sums\cite{pielalattice,ewald,kudinscuseria}. Following this, a number of schemes have been proposed to compute periodic Coulomb integrals using DF approximations\cite{chan-df,riper1,riper4,usvyat1,usvyat2,usvyat_df}. In paper I of this series, we generalise the charge and dipole corrected density fitting approach to local density fitting of MP2 integrals, which we use to obtain DLPNO-MP2 energies subject to BvK boundary conditions.

The complications associated with lattice summation can be removed under the condition that the functions are sufficiently close to the thermodynamic limit, since the summation imposed by the BvK boundary conditions disappears in the limit of an infinite sized supercell. To retain translational invariance, where orbitals at the edge of the simulated supercell are treated identically to those at the centre, it is necessary to account for a larger portion of the material surrounding and including the supercell, which we refer to as the megacell.

In the following sections, we outline the key ideas of our periodic megacell-DLPNO-MP2 approach. First, we discuss the approximation that enables us to remove BvK boundary conditions from our functions, greatly simplifying the evaluation of the ERIs, and outline the relevant boundaries within our simulation cell and the megacell embedding. We then provide details of the generalization of PAOs, OSVs, PNOs and local density fitting, using translational symmetry where appropriate to reduce the computational cost.

\subsection{Megacell embedding}

In the BvK AO basis, the WFs are
\begin{align}
\label{eq:bvkwf}
	\ket{p_{\mathbf{l}}}&=\sum_{\mu_\mathbf{m}}  \ket{\mu_{\mathbf{l}+\mathbf{m}}}C_{\mu_{\mathbf{l}+\mathbf{m}}}^{p_\mathbf{0}} \\
    &=  \sum_{\mu_\mathbf{m}}  \sum_{\mathbf M}^\infty\ket{\mathring{\mu}_{\mathbf{l}+\mathbf{m}+\mathbf{M}}}C_{\mu_{\mathbf{l}+\mathbf{m}}}^{p_\mathbf{0}} ,
\end{align}
where the summation over $\mu_\mathbf{m}$ includes all functions in the supercell and we have used the translational symmetry of the WFs $C_{\mu_{\mathbf{m}}}^{p_\mathbf{l}}=C_{\mu_{\mathbf{l}+\mathbf{m}}}^{p_\mathbf{0}}$, where it is understood that $\mathbf{l}+\mathbf{m}$ is subject to modulo arithmetic within the supercell lattice.
If the WF $\ket{p_{\mathbf{l}}}$ decays to zero within half a supercell extent in all periodic directions, then one can define WFs for the infinite crystal without periodic boundary conditions as
\begin{equation}\label{eq:thermolimit}
	\ket{p_{\mathbf{l}+\mathbf{L}}}=\sum_{\mu_\mathbf{m}}  \ket{\mathring{\mu}_{\mathbf{m}+\mathbf{l}+\mathbf{L}}}C_{\mu_{\mathbf{l}+\mathbf{m}}}^{p_\mathbf{0}}  ,
\end{equation}
where the summation over $\mu_\mathbf{m}$ only extends over one supercell. Provided that the WFs have decayed to zero at the supercell boundary, these WFs form an orthonormal set
\begin{equation}\label{eq:ortho}
    \langle p_{\mathbf{l}}  \vert q_{\mathbf{m}} \rangle = \delta_{pq} \delta_{\mathbf{lm}},
\end{equation}
where the overlap refers to a simple direct-space integral, without periodic boundary conditions.
\begin{figure*}
\includegraphics[scale=0.5]{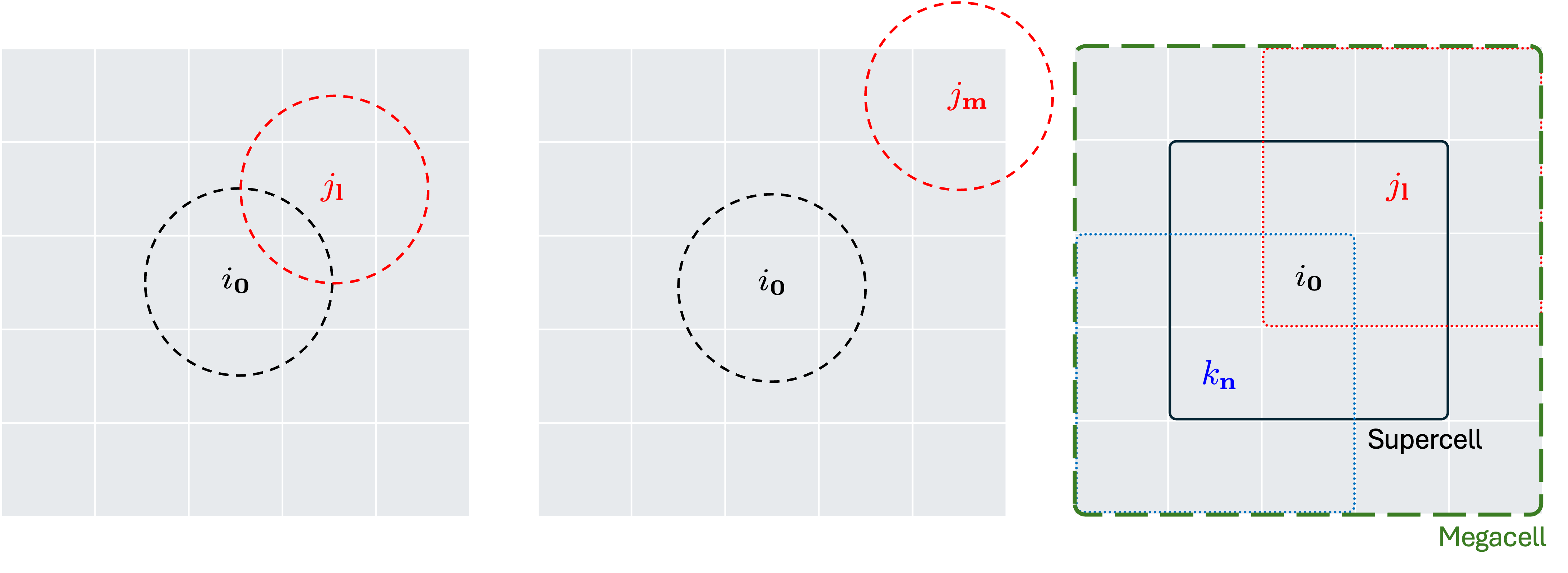}
\caption{\label{fig:ortho}Wannier functions for a two-dimensional system defined by a simulation grid consisting of 5x5 unit cells. Dotted lines represent the extent of the WF, beyond which it has sufficiently decayed. The left subfigure shows a pair of WFs that are fully supported within the simulation grid, the middle subfigure shows a pair of WFs that are not.  The right subfigure defines the boundaries of the supercell and megacell; all WFs in the supercell are properly represented and form a translationally symmetric orthonormal set that can be used for the correlation treatment.}
\end{figure*}

To make use of this simplification in a practical calculation, every orbital used for the correlation treatment must be fully supported within the simulation, so that artifacts due to edge effects are avoided, thereby ensuring full translational symmetry. We therefore embed the \emph{supercell} in a larger \emph{megacell} such that all WFs centered in the supercell are sufficiently decayed at the boundary of the megacell. Only the WFs of the supercell are included in the correlation treatment. This is exemplified in Figure \ref{fig:ortho}, which depicts, in a two-dimensional example, Wannier functions centered in different cells, with their radius of decay.
The megacell boundary adds on half of the extent of the supercell in each lattice vector. Specifically, the megacell lattice extent is given as $k_\mathrm{mega}=2k_\mathrm{super}-1$, where $k_\mathrm{super}$ is the supercell size. We find that well localised occupied WFs often decay rapidly and the orthonormality error is less than one part in a million in our calculations.  The condition that virtual WFs are sufficiently decayed to zero at the supercell boundary is rarely satisfied in practice, and is highly basis set dependent. The DLPNO approach, however, makes use of PAOs instead of virtual WFs, which do decay rapidly to zero, since their extent is directly linked to the occupied orbitals. 


The MP2 residual and energy equations in the Wannier basis (Eq.~\ref{eq:pnomp2res}) can now be employed to correlate all orbitals within the supercell without reference to BvK boundary conditions. Translational symmetry of the integrals and amplitudes remains, but translation extends into the megacell rather than wrapping around the periodic boundary. Coulomb and overlap integral evaluation therefore become identical to that of molecular calculations, without lattice summation. The thermodynamic limit is approached by increasing the size of the supercell to account for all significant pairwise correlations, which requires a commensurate increase in the size of the megacell.

\subsection{Translational symmetry}

The megacell-DLPNO-MP2 method proceeds by first performing a periodic HF calculation in the megacell. Wannier occupied and virtual orbitals are obtained from the Bloch functions and the Fock matrix is transformed to the Wannier basis. The orbitals and Fock elements are then fed into the molecular DLPNO-MP2 code, which applies the PAO-OSV-PNO cascade and local density fitting to compute the correlation energy of the supercell. In contrast to the molecular case, each quantity has the translational symmetry of the crystal lattice. Therefore it is only necessary to compute and store the minimal set of unique quantities. In the following, we discuss how translational symmetry can be used to obtain sub-linear scaling in computational costs.

\subsubsection{PAOs}
PAOs are the projection of the AOs onto the space spanned by the virtual orbitals. The PAOs located in unit cell $\mathbf l$ are translational copies of those in the reference cell
\begin{align}
\vert \tilde \mu_{\mathbf{0}} \rangle &=  \vert \mu_{\mathbf{0}} \rangle - \sum_{i_{\mathbf{l}}} \vert i_{\mathbf{l}} \rangle \langle i_{\mathbf{l}} \vert \mu_{\mathbf{0}} \rangle 
= \sum_{\nu_{\mathbf{m}}} \vert \nu_{\mathbf{m}} \rangle \hat C^{\mu_\mathbf{0}}_{\nu_\mathbf{m}} \\
\vert \tilde \mu_{\mathbf{l}} \rangle &= \sum_{\nu_{\mathbf{m}}} \vert \nu_{\mathbf{m}+\mathbf{l}} \rangle \hat C^{\mu_\mathbf{0}}_{\nu_\mathbf{m}}
\end{align}
where  $\mathring{\mu}_\mathbf{l}=\mu_\mathbf{l}$ in the megacell approach. It is only necessary to compute and store $C^{\mu_\mathbf{0}}_{\nu_\mathbf{lm}} $. Only nearby  $i_\mathbf{l}$ have significant overlap with $\mu_\mathbf{0}$ and the coefficients $\hat C^{\mu_\mathbf{0}}_{\nu_\mathbf{lm}} $ decay rapidly to zero with increasing $\mathbf m$.  
Our approach to forming OSVs employs a numerical Laplace transformation\cite{Almlof_CPL_1991} and we require Laplace transformed PAOs. These are defined as 
\begin{align}
\vert \tilde \mu_{\mathbf{0}}^z \rangle &= \sum_{a_{\mathbf{k}}} \vert a_{\mathbf{k}} \rangle e^{-(\epsilon_{a_\mathbf{k}}-\epsilon_F)t_z} \langle a_{\mathbf{k}} \vert \tilde \mu_{\mathbf{0}} \rangle = \sum_{\nu_{\mathbf{m}}} \vert \nu_{\mathbf{m}} \rangle \hat C^{\mu_\mathbf{0},z}_{\nu_\mathbf{m}} 
\end{align}
where $\epsilon_F$ is the Fermi level and $t_z$ is a Laplace integration grid point and $a_\mathbf{k}$ is a canonical virtual orbital, obtained from the Bloch orbitals of the megacell HF calculation. The Laplace transformed
PAOs $\vert \tilde \mu_{\mathbf{l}}^z \rangle$ are also all translational copies of $\vert \tilde \mu_{\mathbf{0}}^z \rangle$. The computational scaling for forming $\tilde \mu_{\mathbf{0}}$ is $\mathcal{O}(1)$ due to the rapid decay of the overlap $\langle i_{\mathbf{l}} \vert \mu_{\mathbf{0}} \rangle $. This is achieved in practice by using sparse matrix routines. Obtaining similar scaling for forming $\tilde \mu_{\mathbf{0}}^z $ is more challenging due to the canonical virtuals which introduces a scaling of at least $\mathcal{O}(N)$ for this step, but with a low prefactor.

\subsubsection{ERIs}
In the local DF approximation, the ERIs are computed using the robust formula
\begin{align}
    g^{i_{\mathbf 0} j_{\mathbf l}}_{\tilde{a}\tilde{b}}
    &=\sum_{Q_{\mathbf{m}},P_{\mathbf{n}}}(i_{\mathbf 0}\tilde{a} |Q_{\mathbf{m}})(Q_{\mathbf{m}}|P_{\mathbf{n}})^{-1}(P_{\mathbf{n}}|j_{\mathbf l}\tilde{b}),
\end{align}
where $Q_{\mathbf{m}},P_{\mathbf{n}}$ are the subset of auxiliary basis functions local to pair $i_\mathbf{0}j_\mathbf{l}$. The local fitting domain is the set of functions with overlaps  $(i_\mathbf{0}i_\mathbf{0}Q_\mathbf{m}Q_\mathbf{m})$ or $(j_\mathbf{l}j_\mathbf{l}P_\mathbf{n}P_\mathbf{n})$ above a DF threshold, which is linked to the PNO threshold.
We restrict the domains of auxiliary basis functions to within the supercell surrounding the occupied WF, so that the domain for $i_\mathbf{0}j_\mathbf{l}$ does not extend beyond the megacell. The largest PAO domains at the start of the PAO-OSV-PNO cascade are selected on the basis of integral estimates, but here we also apply a restriction to within the supercell surrounding the occupied WF.
The two-index and three-index integrals exhibit translational symmetry 
\begin{align}
     (Q_{\mathbf{m}}|P_{\mathbf{n}}) &= (Q_{\mathbf{0}}|P_{\mathbf{n}-\mathbf{m}}), \\
    (i_{\mathbf{l}}\tilde{\mu}_{\mathbf{n}}|Q_{\mathbf{m}})&=(i_{\mathbf{0}}\tilde{\mu}_{\mathbf{n}-\mathbf{l}}|Q_{\mathbf{m}-\mathbf{l}}), \\
        (i_{\mathbf{l}}\tilde a|Q_{\mathbf{m}})&=(i_{\mathbf{0}}\tilde a |Q_{\mathbf{m}-\mathbf{l}}).
\end{align}
It is only necessary to compute and store the unique integrals where one index is in the reference cell. The computational cost of evaluating $(i_{\mathbf{0}}\tilde{\mu}_{\mathbf{n}}|Q_{\mathbf{m}})$ is asymptotically independent of supercell size once AO integral screening is applied, since the DF and PAO domains are of $\mathcal{O}(1)$ once the supercell is large enough for these domains to saturate. 

\subsubsection{OSVs}
Orbital specific virtuals are PNOs for diagonal pairs $\vert i_\mathbf{l}i_\mathbf{l}\rangle$. OSVs are used as an intermediary for the purpose of accelerating the determination of PNOs, where the PNOs for pair $i_\mathbf{0}j_\mathbf{l}$ are selected from the union of OSVs for $i_\mathbf{0}$ and $j_\mathbf{l}$. It is only necessary to compute and store the OSVs WFs of the reference cell $\vert i_\mathbf{0}\rangle $ since those for $\vert i_{\mathbf{l}}\rangle$ can be obtained from translational symmetry. The procedure follows that for the molecular case, where the OSVs for
orbital $\vert i_{\mathbf{0}}\rangle$ are eigenvectors of an external density matrix for pair 
$\vert i_{\mathbf{0}}i_{\mathbf{0}}\rangle$, which are constructed in a principal domain of 
PAOs $\tilde \mu \in \mathcal{D}_{i_{\mathbf{0}}}$ selected using the greedy algorithm described in Ref.\citenum{tew-principal}.
The external density is computed from first-order amplitudes for diagonal pairs approximated using a numerical Laplace transformation\cite{almlof-laplace,Almlof_CPL_1991}. The integration points and 
weights $w_z$ are determined from the orbital eigenvalues of the supercell in the same way as for molecular calculations. The OSVs for orbital $i_\mathbf{l}$ are related to those of orbital $i_\mathbf{0}$ through
\begin{align}
\vert \tilde{a}_{i_\mathbf{l}}\rangle = \sum_{\tilde{\mu}_\mathbf{m}} \vert \tilde{\mu}_{\mathbf{m}+\mathbf{l}}\rangle C^{\tilde a_{i_\mathbf{0}}}_{\tilde{\mu}_{\mathbf{m}}}
\end{align}
where $C^{\tilde a_{i_\mathbf{0}}}_{\tilde{\mu}_{\mathbf{m}}}$ are the PAO to OSV coeffcients for $i_\mathbf{0}$. In the limit of supercell sizes where the number of DF functions and PAOs above the screening thresholds saturate, the cost of forming the OSVs becomes $\mathcal{O}(1)$ for each $i_\mathbf{0}$ and therefore $\mathcal{O}(1)$ overall.
\subsubsection{Pair Screening}
Once the OSVs are determined, the OSV-SOS-MP2 energy can be used to compute approximate pair energies for the purpose of discarding insignificant pairs and providing an estimate of their contribution to the total energy. The OSV-SOS-MP2 energy does not contain exchange contributions and can be evaluated efficiently using asymmetric DF, which requires only translational copies of the $(i_{\mathbf{0}}\tilde a|Q_{\mathbf{m}})$ integrals, without merging DF domains\cite{tewquasi}.
Only pair estimates contributing to the energy of the reference cell need to be evaluated, following the convention set by Eq.~\ref{eq:emp2wannier},
\begin{align}
E^{i_{\mathbf{0}} j_{\mathbf{m}}}_\text{SOS} = - \!\!\!\!\!\! \sum_{\bar a \in [i_{\mathbf{0}}] \bar b \in [j_{\mathbf{m}}]} 
\frac{g_{\bar a \bar b}^{i_{\mathbf{0}} j_{\mathbf{m}}} g_{\bar a \bar b}^{i_{\mathbf{0}} j_{\mathbf{m}}} }
{f_{i_{\mathbf{0}}}^{i_{\mathbf{0}}} + f_{j_{\mathbf{m}}}^{j_{\mathbf{m}}} - \varepsilon_{\bar a} - \varepsilon_{\bar b}}.
\end{align}
The computational cost of this step is $\mathcal{O}(N)$ if all pairs $i_{\mathbf{0}} j_{\mathbf{m}}$ are considered. In the limit of large supercell sizes, simple prescreening approaches such as distance based cutoffs or dipole approximations can be used to truncate the pair list to obtain $\mathcal{O}(1)$ scaling.

\subsubsection{PNOs}
PNOs for pair $i_\mathbf{0}j_\mathbf{l}$ are formed from the model density given by semi-canonical MP2 amplitudes constructed in the union of the OSVs for $i_\mathbf{0}$ and $j_\mathbf{l}$. Specifically,
\begin{align}
t^{i_\mathbf{0}j_\mathbf{l}}_{\bar a\bar b} &= 
- (\varepsilon_{\bar a} - \varepsilon_{\bar b} - f_{i_\mathbf{0}}^{i_\mathbf{0}} -f_{j_\mathbf{l}}^{j_\mathbf{l}}  )^{-1}
g_{\bar a \bar b}^{i_\mathbf{0} j_\mathbf{l}}
\\
u^{i_\mathbf{0}j_\mathbf{l}}_{\bar a\bar b} &= 2t^{i_\mathbf{0}j_\mathbf{l}}_{\bar a\bar b} - t^{i_\mathbf{0}j_\mathbf{l}}_{\bar b\bar a} \\
D_{\bar a}^{\bar b} &= 2 \sum_{\bar c} ( t^{i_\mathbf{0}j_\mathbf{0}}_{\bar a\bar c} u^{i_\mathbf{0}j_\mathbf{l}}_{\bar b\bar c} + t^{i_\mathbf{0}j_\mathbf{l}}_{\bar c\bar a} u^{i_\mathbf{0}j_\mathbf{l}}_{\bar c\bar b} )
\end{align}
where $\{\bar a\}= \{\tilde{a}_{i_\mathbf{0}}\} \cup \{\tilde{a}_{j_\mathbf{l}}\}$. The PAO and DF domains are also the union of those for $i_\mathbf{0}$ and $j_\mathbf{l}$.  Computing the Coulomb and exchange integrals for each pair is one of the most costly steps in the DLPNO-MP2 method. Translational symmetry reduces the cost of evaluating the density fitting integrals, since it is only necessary to compute $(i_\mathbf{0}\tilde{\mu}_\mathbf{m}|Q_\mathbf{n})$ in the full PAO and DF union space, but the transformation to the pre-PNO space $\bar a$ and integral assembly needs to be preformed for all pairs. This step scales linearly with the size of the supercell up to the point where the number of significant pairs $i_\mathbf{0}j_\mathbf{l}$ within the supercell plateaus.

\begin{figure}
	\includegraphics[scale=0.7]{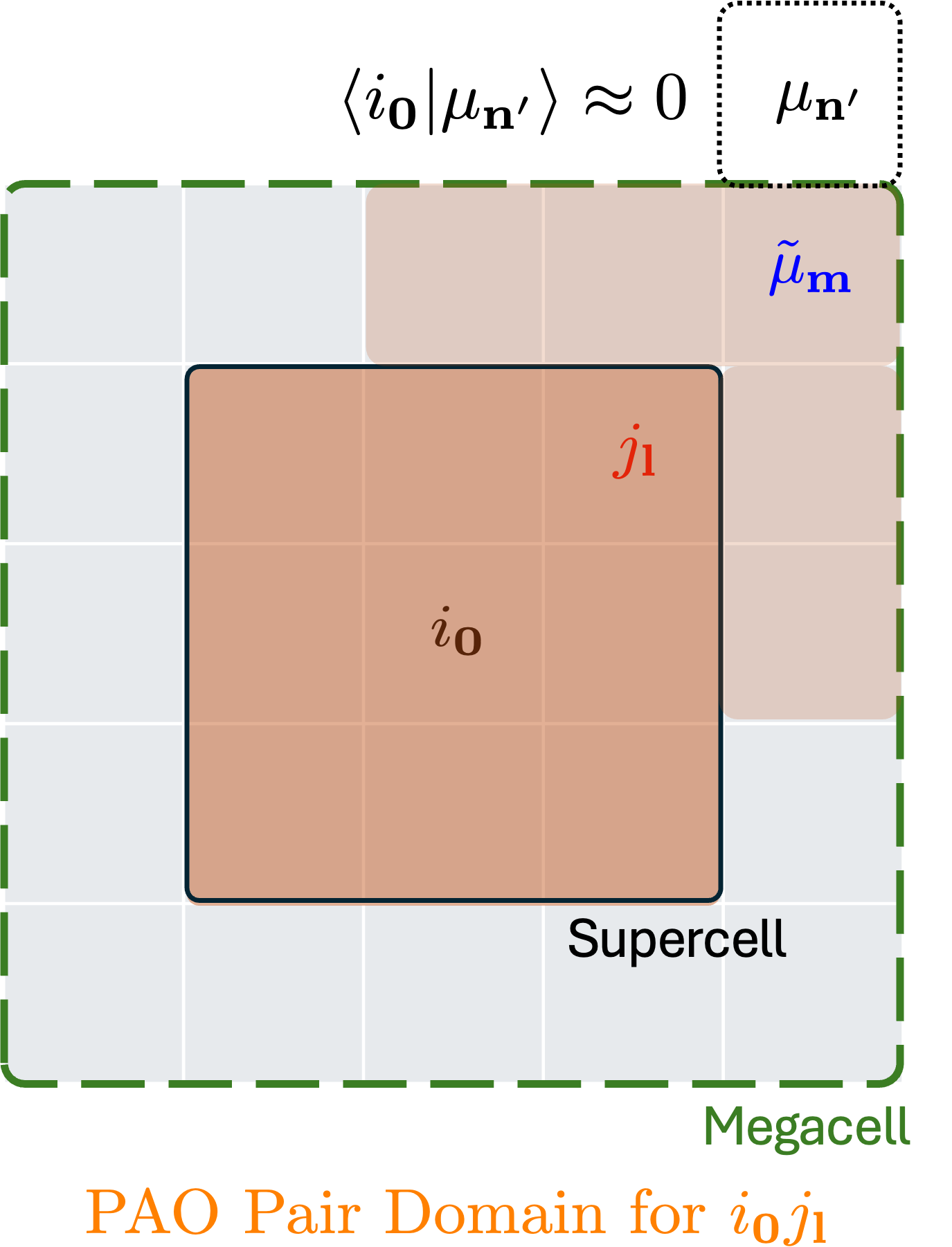}
	\caption{\label{fig:paodomain}The PAO domain for pair $i_{\mathbf{0}}j_{\mathbf{l}}$ spans the entire megacell. The deeper shaded orange region shows the range of the PAOs that are fully supported within the megacell. PAOs within the lighter shaded orange regions, such as $\tilde{\mu}_{\mathbf{m}}$, are missing AOs such as $\mu_{\mathbf{n'}}$ to represent them, but the contributions of these AOs to the PAO DF integrals are negligible due to minimal overlap with $i_\mathbf{0}$.}
\end{figure}

The domain of PAOs for a pair $i_\mathbf{0} j_\mathbf{l}$ touches the boundary of the megacell and the AOs required to form a PAO at the megacell boundary extend beyond the megacell. This is illustrated in Figure \ref{fig:paodomain}  using a two-dimensional example. However, the contribution to $(i_\mathbf{0}\tilde{\mu}_\mathbf{m}|Q_\mathbf{l})$ from an AO outisde of the megacell $(i_\mathbf{0}\mu_{\mathbf{n}'}|Q_\mathbf{l})$ is completely negligible since the overlap of those AOs with $i_\mathbf{0}$ vanishes and the contribution would anyway have been screened out. Since the corresponding integrals for $j_\mathbf{l}$ are translational copies from the reference cell, the robustness in the integrals is preserved for all occupied orbitals in the supercell.

\subsubsection{DLPNO-MP2 Residuals and Energy}
The MP2 energy under the semi-canonical approximation (DLPNO-sc-MP2) can be computed as a byproduct of forming PNOs,
\begin{align}
E^{i_\mathbf{0} j_\mathbf{l}}_\text{SC} = \sum_{\bar a \bar b \in [i_\mathbf{0}j_\mathbf{l}]} 
u_{\bar a \bar b}^{i_\mathbf{0} j_\mathbf{l}} g_{\bar a \bar b}^{i_\mathbf{0} j_\mathbf{l}}.
\end{align}
The difference between the PNO-sc-MP2 energy before and after truncation of the PNO space 
to $\{\tilde a_{i_\mathbf{0}j_\mathbf{l}}\}$ provides an estimate for the contribution of the discarded
PNOs to the total energy. The sc-MP2 energy neglects the coupling due to the off-diagonal Fock elements. The final DLPNO-MP2 amplitude equations (Eq.~\ref{eq:pnomp2res}) are solved iteratively in the space of retained PNOs. The coupling introduces amplitudes and overlaps for pairs $\{k_{\mathbf{m}}j_{\mathbf{l}}\}$ where neither index is in the reference cell, but these are translational copies of a pair with one index in the reference cell, and the residual equations become
\begin{align}
0 =& 
g^{i_{\mathbf 0} j_{\mathbf l}}_{\tilde a \tilde b} + 
(\varepsilon_{\tilde a} + \varepsilon_{\tilde b}) t^{i_{\mathbf 0} j_{\mathbf l}}_{\tilde a \tilde b} \nonumber \\
& - \sum_{k_{\mathbf{m-l}}} \sum_{\tilde c \tilde d \in [ k_{{\mathbf{m-l}}}j_{{\mathbf 0}} ] }
S_{\tilde a, i_{\mathbf{-l}} j_{\mathbf 0}}^{\tilde c, k_{\mathbf{m-l}} j_{\mathbf{0}}}
S_{\tilde b, i_{\mathbf{-l}} j_{\mathbf 0}}^{\tilde d, k_{\mathbf{m-l}} j_{\mathbf{0}}}
t^{ k_{\mathbf{m-l}} j_{\mathbf 0}}_{\tilde c \tilde d} f_{k_{\mathbf m}}^{i_{\mathbf 0}}  \nonumber \\
& - \sum_{k_{\mathbf m}} \sum_{\tilde c \tilde d \in [ i_{{\mathbf 0}}k_{{\mathbf m}} ] }
S_{\tilde a, i_{\mathbf 0} j_{\mathbf l}}^{\tilde c, i_{\mathbf 0} k_{\mathbf{m}}}
S_{\tilde b, i_{\mathbf 0} j_{\mathbf l}}^{\tilde d, i_{\mathbf 0} k_{\mathbf{m}}}
t^{ i_{\mathbf 0} k_{\mathbf m}}_{\tilde c \tilde d} f_{k_{\mathbf m - \mathbf l}}^{j_{\mathbf 0}}.
\end{align}
The final DLPNO-MP2 correlation energy per unit cell is given by
\begin{align}
E_\mathrm{corr}= \frac{2}{1+\delta_{i_{{\mathbf 0}}j_{{\mathbf l}}}} \sum_{i_{{\mathbf 0}}\le j_{{\mathbf l}}} \sum_{\tilde a \tilde b \in [ i_{{\mathbf 0}}j_{{\mathbf l}} ] }
 (2g^{i_{\mathbf{0}}j_{\mathbf{l}}}_{\tilde a \tilde b} - g^{i_{\mathbf{0}}j_{\mathbf{l}}}_{\tilde b \tilde a})
t^{i_{\mathbf{0}}j_{\mathbf{l}}}_{\tilde a \tilde b} + \Delta,
\label{eq:emp2delta}
\end{align}
where $\Delta$ is the correction term composed of the energy estimates for the discarded pairs and PNOs. 

\subsection{Wannier Function Localization}
\label{sec:wf}
The efficacy of the megacell embedding scheme is contingent on having well-localised occupied WFs that decay sufficiently rapidly. The Foster--Boy's method\cite{FB-1,FB-2} has seen widespread use for obtaining well-localised WFs of plane-wave treatments of crystalline systems \cite{Marzari-Vanderbilt,Wannier90}, but it is more convenient in LCAO approaches to use Pipek--Mezey schemes\cite{Jonsson,Clement,schreder-luber}.

In this work, we choose to localize the WFs by maximising the fourth moment of the Pipek--Mezey metric using  partial charges obtained from Knizia's IAOs\cite{knizia2013}
\begin{equation}{\label{PMmetric}}
\braket{O}_{\textbf{PM}}=\sum_{\mathbf{l},A,i}|Q_i^{A_{\mathbf{l}}}|^4=\sum_{\mathbf{l},A,i}\bra{i_{\mathbf{0}}}\hat{P}_{A_{\mathbf{l}}}^{\mathrm{IAO}}\ket{i_{\mathbf{0}}}^4, 
\end{equation}
where the projector $\hat{P}_{A_{\mathbf{l}}}^{\mathrm{IAO}}$ involves a restricted summation over the IAOs belonging to the atom $A$ in cell $\mathbf{l}$. Localization of the WFs is greatly aided by our diabatic Wannierisation procedure, which serves as an initial guess by fixing the gauge the of the orbitals to vary smoothly with the Bloch functions at the $\Gamma$- point. We refer the reader to Ref \onlinecite{zhuwannier} for the implementation details of this approach. The use of IAOs rather than the original Muliken charges removes the problem of strong basis set dependence, and the use of the fourth moment penalises the tails of the WFs, leading to a more rapid decay.

\section{Computational Details}
Megacell domain-based local PNO-MP2 (megacell-DLPNO-MP2) has been implemented in a developmental version of the TURBOMOLE\cite{Turbomole} program, within the \verb;pnoccsd; module\cite{Schmitz01092013,tew-principal}. Periodic LCAO-based Hartree--Fock calculations using $\mathbf k$-point sampling have recently become available, in the \verb;riper; module\cite{riper1,riper2,riper3,riper4,riper5}, the output of which provides the HF Bloch functions and band energies required for MP2. All HF calculations utilized the universal Coulomb-fitting auxiliary basis sets\cite{universalauxbas} for the RI-J approximation. In this work we adopt the Monkhorst-Pack grid for our $\mathbf k$-point grid\cite{kmesh}. The Wannier function localization procedure has been implemented in a developmental version of the \verb;riper; module \cite{zhuwannier}. We benchmark and test the performance of the periodic megacell-DLPNO-MP2 scheme using a range of one-, two- and three-dimensional systems including linear C$_2$HF polymers,
hexagonal Boron Nitride sheets, two rocksalt structures (LiH, MgO), one diamond cubic structure (Si), as well a hexagonal polymorph of ice. All calculations employed the pob-TZVP\cite{pob-TZVP} orbital basis sets. The density fitting approximation used in the MP2 calculations employed the def2-TZVP auxiliary basis sets\cite{def2tzvp}. All calculations were run on a single node (Intel(R) Xeon(R) Gold 6248R CPU) with a maximum RAM limit of 386\,GB and 1.8\,TB disk, with OMP parallelization. Lattice constants and computed energies for all calculations are collected in the supplementary information.

The computational scaling of the algorithm is expected to tend to $\mathcal{O}(1)$, as supercell size approaches the thermodynamic limit, and the local domain approximation and the pair screening combine such that the number of $i_\mathbf{0}j_{\mathbf{l}}$ pairs scales to a constant. We probe this scaling behavior numerically in the results section. 

\section{Results}

\begin{figure}[t]
    \centering
	\includegraphics[scale=0.43]{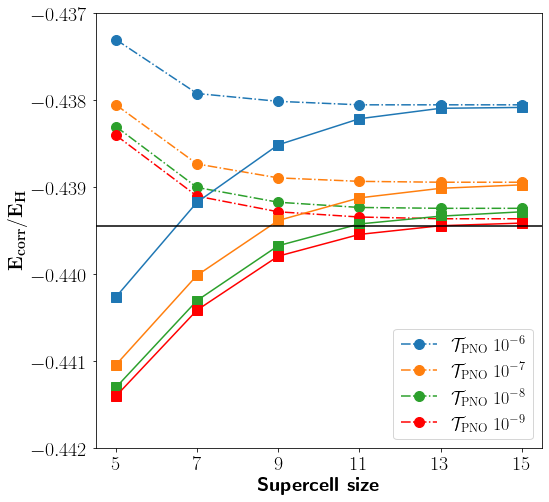}
	\caption{\label{fig:c2hftol} Megacell (circles) and BvK (squares) DLPNO-MP2 correlation energies of $\mathrm{C_2HF}$ as a function of 1D $k$-mesh and $\mathcal{T}_\text{PNO}$. The horizontal line indicates the canonical thermodynamic limit\cite{bvk}.}
\end{figure}

\begin{figure}[t]
    \centering
	\includegraphics[scale=0.43]{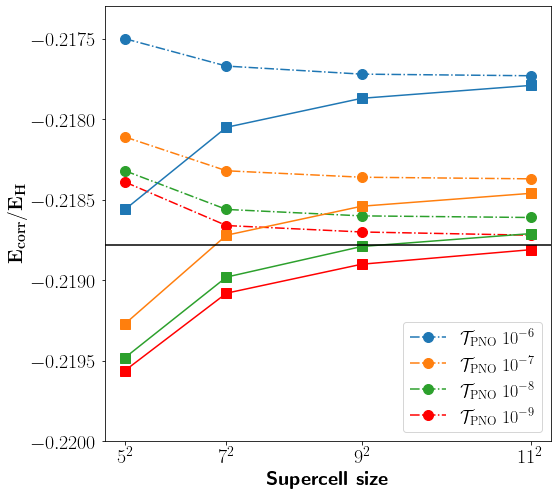}
	\caption{\label{fig:bntol} Megacell (circles) and BvK (squares) DLPNO-MP2 correlation energies of BN as a function of 2D $k$-mesh and $\mathcal{T}_\text{PNO}$. The horizontal line indicates the canonical thermodynamic limit\cite{bvk}.}
\end{figure}

\subsection{PNO and $k$-mesh convergence}

In Paper I we report DLPNO-MP2 and canonical MP2 correlation energies at the thermodynamic limit for the $\mathrm{C_2HF}$ polymer chain and the
Boron Nitride sheet using the pob-TZVP AO basis set and the def2-TZVP Coulomb fitting basis set. These values were obtained from a series of molecular fragment calculations by removing edge effects, and provide benchmarks
against which to assess the rate of convergence of the megacell-DLPNO-MP2 correlation energy to the thermodynamic limit as a function of supercell size.

Figure \ref{fig:c2hftol} presents correlation energies computed using the megacell and BvK-DLPNO-MP2 methods for a one-dimensional $\mathrm{C_2HF}$ chain, with increasing supercell size, at varying ${\mathcal T}_\text{PNO}$ thresholds. Figure \ref{fig:bntol} presents the equivalent data for the
2D sheets of Boron Nitride. In both examples, the BvK and megacell schemes converge to the same energy at the thermodynamic limit for each PNO threshold and these values coincide with the benchmarks obtained from the molecular fragment calculations. This remarkable agreement between the three different schemes at the thermodynamic limit for each PNO threshold underlines the consistency in the PNO local correlation approximation and the stability of the PAO-OSV-PNO cascade. 

Comparing the convergence of the BvK and Megacell approaches to the thermodynamic limit, we find that BvK correlation energies converge from below, whilst the megacell values converge from above. The rate of convergence for the megacell approach is faster than that of BvK, which is most likely because the HF orbitals and band energies for the megacell are closer to the thermodynamic limit than those of the BvK supercell calculation.

Paper I also reports BvK-DLPNO-MP2 correlation energies for three-dimensional Lithium Hydride (LiH) and Magnesium Oxide (MgO) using $\mathcal{T}_\text{PNO} = 10^{-6},10^{-7},10^{-8}$ and a series of $k$-meshes. Table \ref{table:mgolihbvkmega} compares the megacell- and BvK-DLPNO-MP2 correlation energies for these systems. For LiH, agreement to better than $0.1$ millihartree is already achieved for a $5^3$ supercell, at each ${\mathcal T}_\text{PNO}$ value. 
For MgO, the rapid convergence of the two schemes supercells is also evident, with differences between megacell and BvK decreasing by an order of magnitude from $3^3$ to $5^3$. In contrast to the earlier one- and two-dimensional examples, both the megacell and BvK schemes appear to converge to the thermodynamic limit from above. Further comparison of the asymptotic behavior of both approaches will be the subject of future work.

\begin{table}  
\centering
\caption{MP2 correlation energy comparison between megacell- and BvK-DLPNO-MP2 implementations, varying PNO truncation threshold and supercell size.}
\label{table:mgolihbvkmega}
\begin{tabular}{|l|l|cc|cc|}
\hline
          System & ${\mathcal T}_\text{PNO}$ & mega & BvK & mega & BvK\\
\hline
         $k_{\mathrm{super}}$  && \multicolumn{2}{c|}{$5^3$} & \multicolumn{2}{c|}{$7^3$} \\

\hline
LiH     & 6   &-0.03063&-0.03066& -0.03083 & -0.03083 \\
 & 7 &-0.03070&-0.03075& -0.03091 & -0.03093\\
 & 8 &-0.03073&-0.03078& -0.03096 & -0.03096 \\
\hline
$k_{\mathrm{super}}$  && \multicolumn{2}{c|}{$3^3$} & \multicolumn{2}{c|}{$5^3$} \\
\hline
MgO    & 6   &-0.19747    & -0.19631   &-0.19868 & -0.19853 \\
 & 7 & -0.19765 &-0.19651& -0.19897 & -0.19882\\
 & 8 & -0.19774 &-0.19660& -0.19908 & -0.19902\\
\hline
\end{tabular}
\end{table}

\begin{figure}[h]
    \centering
	\includegraphics[scale=0.4]{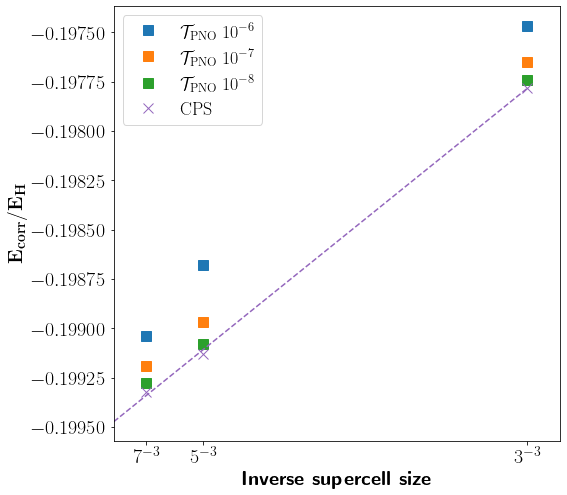}
	\caption{\label{fig:mgocpstdl} MP2 correlation energies for MgO, obtained with increasing PNO truncation thresholds and larger supercells. Extrapolations to the complete PNO space limit for each supercell size are given, and an inverse volume extrapolation to the thermodynamic limit is plotted.}
\end{figure}

Finally, to assess the degree to which megacell-DLPNO-MP2 is able to accurately estimate the canonical and thermodynamic limit, in Figure \ref{fig:mgocpstdl} we plot the MP2 correlation energies of MgO as a function of inverse supercell size (volume), at varying PNO truncation thresholds. For each supercell size, the complete PNO space (CPS) limit is obtained through a square root extrapolation\cite{keshaextrpolate,keshadamyan} of the PNO truncation threshold, motivated by the observation that the largest discarded amplitude is proportional to the square of the threshold. An inverse volume extrapolation\cite{boothfciqmcsolid,keliaofinite} is then performed using the CPS values to obtain a canonical and thermodynamic limit value for the MP2 correlation energy per cell. To a very good approximation, the megacell-DLPNO-MP2 values lie on the expected line following the inverse volume convergence. The consistent behaviour of the megacell-DLPNO-MP2 scheme thus makes it possible to extrapolate to both the CPS and thermodynamic limits.

\subsection{Computational Scaling}

For molecules, the combined effect of pair screening, integral screening, local density fitting and PNO compression leads to an asymptotically linear-scaling DLPNO-MP2 method with respect to system size.\cite{Schmitz01092013,tewhattigpnomp2,turbomoletodaytom} In essence the complexity for correlating each orbital $i$ is asymptotically $\mathcal{O}(1)$ since correlation is short-ranged and linear scaling results from there being $\mathcal{O}(N)$ orbitals in a molecule. For periodic systems, this equates to an asymptotic scaling that is linear in the number of orbitals in a unit cell, which is $\mathcal{O}(1)$. This formal scaling of computational effort is only observed in practice if appropriate local approximations are applied to every step in the program workflow, and the system size is sufficiently large for the locality savings to take effect. In the following, we report the observed scaling of our current implementation for 2D Boron Nitride and 3D Lithium Hydride systems in realistic calculations. All CPU timings were evaluated on a single OMP thread.

\begin{figure*}[ht]
\centering
\includegraphics[scale=0.45]{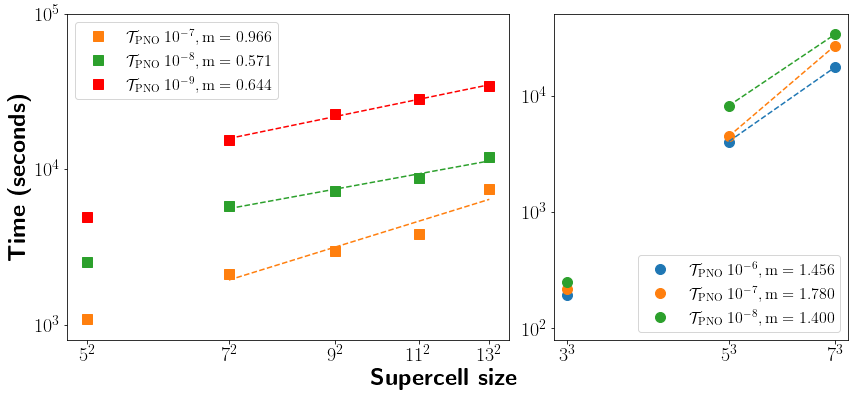}
\caption{Log scaled CPU times for two-dimensional Boron Nitride (left), and three-dimensional Lithium Hydride (right), as a function of log scaled supercell size and PNO truncation threshold. Slope values for the trend lines of the largest supercell sizes as given as $\mathrm{m= [slope value]}$}\label{fig:bnlihtottimings}
\end{figure*}

\begin{figure*}[t]
\centering
\includegraphics[scale=0.45]{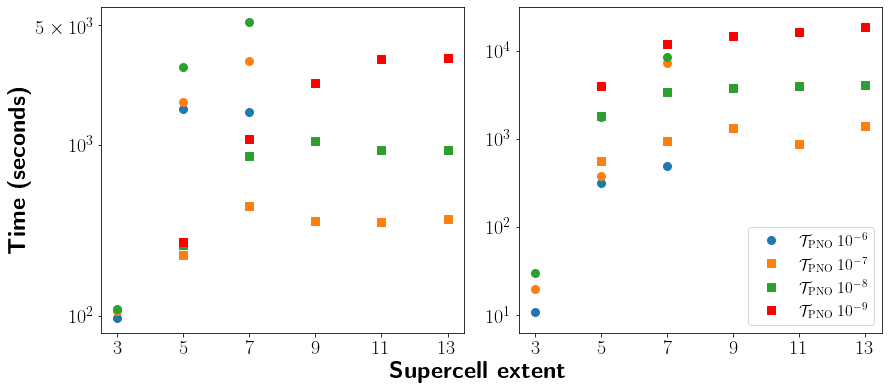}
\caption{Log scaled CPU times for two key subroutines in the megacell-DLPNO-MP2 method, as a function of the supercell extent in each dimension, and PNO truncation threshold. The left panel shows the generation of the three index density-fitting integrals, whilst the right panel presents the  construction of the approximate external pair density, and the subsequent diagonalization to construct the PNOs. Circle markers denote timings from three-dimensional Lithium Hydride. Square markers denote timings from two-dimensional Boron Nitride.}\label{fig:bnlihpdens3timings}
\end{figure*}

In Figure \ref{fig:bnlihtottimings}, the total CPU times for Boron Nitride (left) and Lithium Hydride (right) are plotted, at increasing supercell sizes. Lines of best fit of the log-scaled graphs, discarding the smallest supercell size, are also presented, and the gradients are shown. For BN, the fitted lines demonstrate overall linear and sub-linear CPU scaling in this size regime. This is a direct consequence of the combination of the local domains, pair screening and translational invariance savings. Whilst the linear and sub-linear scaling from our BN results are encouraging, the theoretical $\mathcal{O}(1)$ scaling is not attained, even though the interaction pair list $i_\mathbf{0}j_{\mathbf{l}}$ has saturated. We attribute this to several subroutines that have not yet been fully adapted from the molecular scheme to leverage translational invariance. Particularly for $\mathcal{T}_\text{PNO}=10^{-7}$, these routines, which were not considered to have expensive scaling or prefactors earlier in the code development, now have a measurable impact of the overall cost.

The CPU scaling for LiH currently shows only sub-quadratic cost at the largest two supercell sizes, for all $\mathcal{T}_\text{PNO}$ values. This is partly due to the aforementioned unadapted routines, which have a greater impact for larger three-dimensional systems. More significantly, however, the largest supercells employed for LiH have not yet entered the regime in which the interaction sphere of $i_\mathbf{0}j_{\mathbf{l}}$ pairs becomes fully saturated. A transition to sub-linear scaling at larger supercell sizes is expected, but we have not been able to demonstrate this due to the limitations of our available hardware.

The two most expensive steps in the megacell-DLPNO-MP2 method are the evaluation of the three-index integrals $(\mu_\mathbf{m} i_\mathbf{0} | Q_\mathbf{l})$ and the formation of the PNOs, which involves the construction of the ERIs in the pre-PNO basis $(\bar{a}i_\mathbf{0} | \bar{b}j_\mathbf{l})$. The individual CPU timings for these key steps in megacell-DLPNO-MP2 are presented in Figure \ref{fig:bnlihpdens3timings}, for both BN and LiH. The log-scaled times are plotted against the supercell extent in each dimension, to enable comparison of the field of interaction between two-dimensional BN and three-dimensional LiH. 

The left panel displays the CPU times for the evaluation of the three-index integrals. For the 2D BN example, the observed CPU cost tends to a constant, reflecting the correct $\mathcal{O}(1)$ scaling due to the saturation of the density fitting domains and efficacy of integral screening. The DF threshold is linked to the PNO threshold and saturation of the DF domain requires larger supercells for tighter PNO thresholds. The 3D LiH example appears to follow a similar trend, but the supercell diameter of 7 unit cells is not sufficiently large for the DF domain to saturate for the asymptotic scaling to manifest. The right panel displays the CPU times for the subroutine where the PNOs are formed. For BN, the observed timings again tend towards $\mathcal{O}(1)$ scaling after a diameter of 9 unit cells. Although the LiH data follows a similar trend for $\mathcal{T}_\text{PNO}=10^{-6}$ and $10^{-8}$, the timings for $10^{-7}$ scales linearly in this regime. This is due to sub-optimal I/O batching that has not yet been adjusted from the molecular code to account for reduced pair list. Although the code can already be used to computed large supercells at tight PNO thresholds, improved performance can be found through further optimisation.

\begin{figure*}[ht!]
\centering
	\includegraphics[scale=0.4]{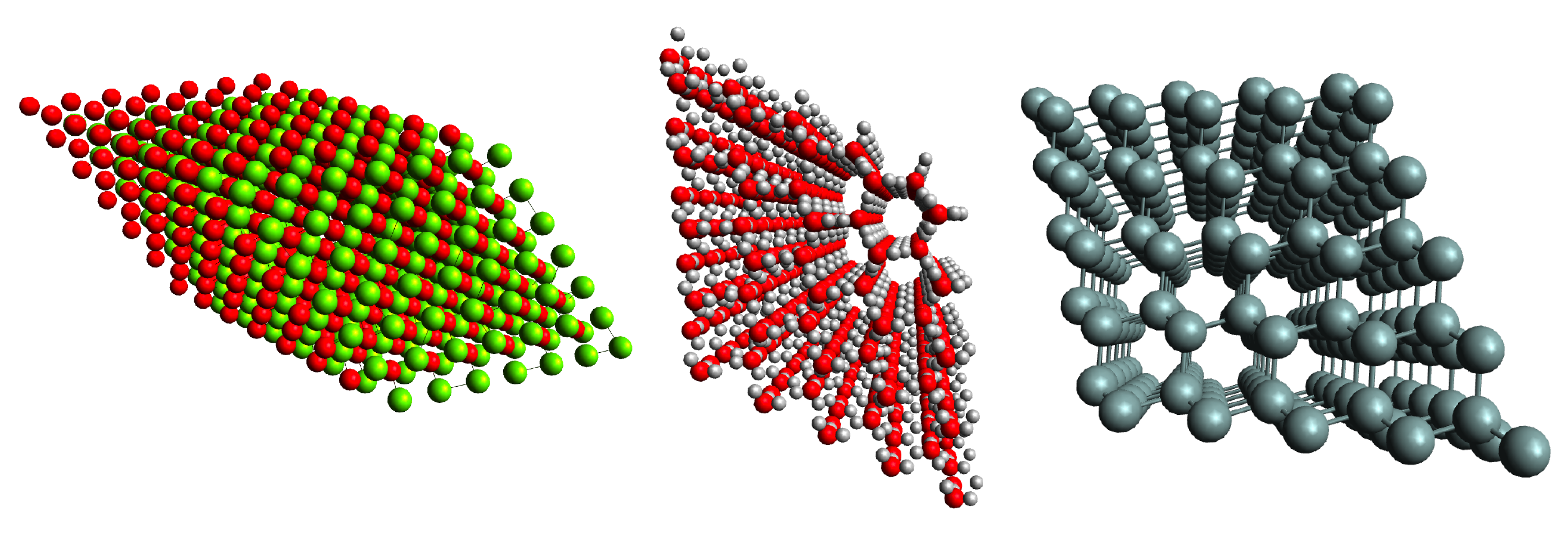}
	\caption{\label{fig:3d} Supercells of the largest three-dimensional systems explored in this contribution. From left to right: 7x7x7 Magnesium Oxide (rock salt), 5x5x5 Hexagonal Ice, 5x5x5 Silicon (diamond cubic). The number of basis functions spanned within each supercell is 12691, 15000 and 5500, respectively.}
\end{figure*}

\subsection{Applications}

To demonstrate the current scope and performance of the megacell-DLPNO-MP2 approach we apply our method to simulate large supercells of three-dimensional materials. Two rocksalt structures (LiH, MgO), one diamond cubic structure (Si), as well a hexagonal polymorph of ice, are probed. The pob-TZVP orbital basis sets are used. Table \ref{tab:3dsystems} presents MP2 correlation energies for the valence bands for each of these systems, at varying PNO truncation thresholds. Canonical pob-TZVP energies are estimated from extrapolation to the complete PNO, and inverse volume extrapolations are applied to obtain the canonical and thermodynamic limit estimates, $E_{\mathrm{corr,TDL}}$. 

The supercells for our three largest calculations are displayed in Figure \ref{fig:3d}. The largest, a $5^3$ supercell of hexagonal ice, contains 15000 basis functions. With the exception of Si, the MP2 correlation energies obtained from the largest supercell and tightest  $\mathcal{T}_\text{PNO}$ threshold are all within 0.2 millihartree of the estimated $E_{\mathrm{corr,TDL}}$ value, demonstrating the extent to which the megacell-DLPNO-MP2 can sample the thermodynamic limit. All calculations were completed within two days on our machines, using OMP parallelisation over 48 threads. We are currently prevented from performing larger calculations due to some memory and disk bottlenecks of our pilot implementation and the hardware we have access to. Nevertheless, the accuracy and efficiency of our approach is encouraging and we anticipate that
that megacell-DLPNO-MP2 method can be readily applied to obtain properties such as lattice constants, cohesive energies, chemisorption and physisorption on surfaces and other ground state properties of non-conducting systems.

\begin{table*}[ht!]  
\centering
\caption{Survey of three-dimensional materials probed using megacell-DLPNO-MP2, given by PNO truncation threshold and largest supercell size employed. Hartree-Fock energies are provided for each supercell. Only valence bands were included in the correlated treatment. }
\label{tab:3dsystems}
\begin{tabular}{|c|c|cc|cc|cc|c|}
\hline
         $k_{\mathrm{super}}$  && \multicolumn{2}{c|}{$3^3$} & \multicolumn{2}{c|}{$5^3$} & \multicolumn{2}{c|}{$7^3$}&  \\
          System & ${\mathcal T}_\text{PNO}$ & $E_{\mathrm{corr}}$ & $E_{\mathrm{HF}}$ & $E_{\mathrm{corr}}$ & $E_{\mathrm{HF}}$ & $E_{\mathrm{corr}}$ & $E_{\mathrm{HF}}$ & $E_{\mathrm{corr,TDL}}$\\
\hline
LiH     & 6   &-0.02934&-8.06060&-0.03063&-8.06059& -0.03083 & -8.06059 & \\
 & 7 &-0.02938&&-0.03070&& -0.03091 &&\\
 & 8 &-0.02940&&-0.03073&& -0.03096 &&\\
 & CPS &-0.02941&&-0.03074&&-0.03098&&-0.03112\\
\hline
MgO    & 6   &-0.19747    &-274.68452   &-0.19868 &-274.68455 & -0.19904 & -274.68455 &  \\
 & 7 & -0.19765 && -0.19897 && -0.19919 && \\
 & 8 & -0.19774 && -0.19908 && -0.19928 && \\
 & CPS &-0.19778&&-0.19913&&-0.19932&&-0.19947\\
\hline
Si     & 6   &-0.16401    &-577.84603 &-0.17394 & -577.84720 &&  &  \\
 & 7 & -0.16507 && -0.17637 &&  && \\
 & 8 & -0.16542 && -0.17703 && && \\
 & CPS &-0.16558&&-0.17734&&&&-0.18057\\
\hline
ice Ih     & 6 &-0.74541&-304.26198&-0.74620&-304.26198&&&  \\
 & 7 &-0.74677&&-0.74738&&&&\\
 & 8 &-0.74727&&-0.74781&&&&\\
 & CPS &-0.74750&&-0.74801&&&&-0.74815\\
\hline
\end{tabular}
\end{table*}

\section{Conclusions}

Wavefunction-based many-body correlation theory provides a systematically improvable hierarchy of methods that can be used to independently verify and benchmark the accuracy of density functional predictions of electronic energies and properties of molecules and materials. In paper I of this series, we presented the periodic generalisation of LCAO-based DLPNO-MP2 theory using Born--von K{\'a}rm{\'a}n periodic boundary conditions. DLPNO local correlation approximations eliminate the steep scaling of computational costs with system size, which is essential when aiming to perform correlated wavefunction calculations using large simulation cells to approach the thermodynamic limit. The goal of our work is to develop the corresponding theory for periodic systems, where we aim to leverage large parts of the existing efficient molecular implementation in the Turbomole program package.

In paper I, we showed through careful benchmarking that the PNO approximations generalise cleanly to the periodic setting and that entirely consistent approximations are made in molecules and materials. However, the BvK boundary conditions introduce expensive lattice summations in the integral evaluation step and require chargeless density fitting with suface dipole corrections, which make it difficult to fully exploit translational symmetry to reduce costs. 

In this contribution, we present an alternative strategy that does not impose BvK
boundary conditions on the correlation treatment. Instead, a supercell is embedded in a larger megacell and periodicity is enforced by imposing translational symmetry on all Hamiltonian matrix elements and wavefunction parameters. Under the assumption that the megacell is effectively infinite, the lattice summation for the integrals is removed and replaced by an infinite sum over pair energies. This has several advantages: since there is no lattice summation, the integrals are less expensive; the integrals are rigorously translationally invariant, so only the minimal unique set need to be computed and stored; the $r^{-6}$ decay of the pair energies is more rapid than the $r^{-3}$ decay of the integrals. The primary disadvantage is that a HF calculation on a very large megacell is required, and the number of basis functions involved when transforming to the Wannier basis presents challenging memory bottlenecks. 

Our calculations show that the PNO approximations underpinning the BvK-DLPNO-MP2 and Megacell-DLPNO-MP2 methods are entirely consistent and that for a given PNO threshold both methods converge to the same thermodynamic limit. Moreover, this consistency extends to the molecular case and it is therefore possible to straightforwardly combine results from molecular and periodic DLPNO calculations when studying molecular insertion in porous solids, or surface adsorption processes. The smooth convergence with PNO threshold to the canonical correlation energy, observed for molecular calculations, is also observed here, and CPS extrapolation to the canonical limit is equally applicable.

Our implementation in the Turbomole package exploits the translational symmetry of the ERIs, Fock matrices, orbital coefficients, overlaps and correlation amplitudes to achieve sub-linear scaling of computational cost with system size. In essence the complexity for correlating each orbital $i$ is asymptotically $\mathcal{O}(1)$ since correlation is short-ranged. For periodic systems, this equates to an asymptotic scaling that is linear in the number of orbitals in a unit cell, which is $\mathcal{O}(1)$.

We have used the Megacell-DLPNO-MP2 method to compute correlation energies at the thermodynamic limit for a series of 1D, 2D and 3D examples, including a 7$\times$7$\times$7 supercell of MgO in a pob-TZVP basis with 12691 basis functions, and a 5$\times$5$\times$5 supercell of hexagonal ice, with 15000 basis functions. The Megacell-DLPNO-MP2 method presented here provides the foundation for efficient periodic implementations of DLPNO-CCSD(T) and DLPNO-CC3 theory for excited states.

\section*{Acknowledgments}
We express our sincere thanks to Dr Denis Usvyat for providing benchmark thermodynamic limit MP2 energies computed using the CRYSCOR scheme.
Financial support for AZ from the University of Oxford and Turbomole GmbH is gratefully acknowledged. AN gratefully acknowledges funding through a Walter Benjamin Fellowship by the Deutsche Forschungsgemeinschaft (DFG, German Research Foundation) -- 517466522.
PK gratefully acknowledges funding through the Institute for the Promotion of Teaching Science and Technology.
\bibliography{refs}


\end{document}